\documentclass[11pt]{article}

\usepackage{graphics}
\usepackage{amsmath,amssymb,amsthm}
\usepackage[latin1]{inputenc}
\usepackage[english]{babel}
\usepackage{comment}
\usepackage{graphicx}

\usepackage{tikz,overpic}

\DeclareMathOperator*{\Tr}{Tr}

\DeclareMathOperator*{\supp}{supp}

\DeclareMathOperator*{\RE}{Re}

\renewcommand{\Re}{\RE}

\newtheorem{theorem}{Theorem}

\newtheorem{proposition}[theorem]{Proposition}

\theoremstyle{definition}

\theoremstyle{remark}

\numberwithin{equation}{section}
\numberwithin{figure}{section}

    \title{Asymptotic analysis of the  two matrix model
     with a quartic potential}

\author{
\ Maurice Duits, \footnote{Department of Mathematics, Royal
Institute of Technology (KTH), Lindstedts\-vagen~25, SE-10044
Stockholm, Sweden, email: duits@kth.se. Supported by the grant KAW
2010.0063 from the Knut and Alice Wallenberg Foundation.} \, \ Arno
B.J.  Kuijlaars, \footnote{Department of Mathematics, Katholieke
Universiteit Leuven, Celestijnenlaan 200 B, 3001 Leuven, Belgium,
email: arno.kuijlaars@wis.kuleuven.be. Supported by FWO grants G.0427.09 and G.0641.11,
K.U. Leuven research grants OT/08/33 and OT/12/73, and
research project MTM2011-28952-C02-01 of the Spanish Ministry of Science and Innovation.}
\, and \, Man Yue Mo
\footnote{Department of Mathematics, University of Bristol, Bristol
BS8 1TW, UK, email: m.mo@bristol.ac.uk. Support by the EPSRC grant
EP/G019843/1.} }
\date{}

\begin{document}
\maketitle
\begin{abstract}
We give a summary of the recent progress made by the authors and collaborators 
on the asymptotic analysis of the two matrix model with a quartic potential.
The paper also contains  a list of open problems.
\end{abstract}

\section{Two matrix model: introduction}

The Hermitian two matrix model is the probability measure
\begin{equation} \label{TwoMM}
    \frac{1}{Z_n} e^{-n \Tr(V(M_1) + W(M_2) - \tau M_1 M_2)} \, dM_1 dM_2
    \end{equation}
defined on pairs $(M_1, M_2)$ of $n \times n$ Hermitian matrices. Here $V$
and $W$ are two polynomial potentials, $\tau \neq 0$ is a coupling constant,
and
\[ Z_n = \int  e^{-n \Tr(V(M_1) + W(M_2) - \tau M_1 M_2)} \, dM_1 dM_2 \]
is a normalization constant in order to make \eqref{TwoMM} a probability measure.

In recent works of the authors and collaborators \cite{DGK,DK,DKM,Mo} the model was studied with the
aim to gain understanding in the limiting behavior of the eigenvalues
of $M_1$ as $n \to \infty$, and to find and describe new types of critical
behaviors.

The results should be compared with the well known results for the Hermitian
one matrix model
\begin{equation} \label{OneMM}
    \frac{1}{Z_n} e^{-n \Tr(V(M))} \, dM,
    \end{equation}
which we briefly summarize here.
The eigenvalues of the random matrix $M$ from \eqref{OneMM} have the explicit joint p.d.f.
\[ \frac{1}{\tilde{Z}_n} \prod_{j < k} (x_k - x_j)^2 \prod_{j=1}^n e^{-n V(x_j)}, \]
which yields that the eigenvalues are a determinantal point
process with correlation kernel
\[ K_n(x,y) = \sqrt{e^{-n V(x)}} \sqrt{e^{-nV(y)}} \sum_{k=0}^{n-1} p_{k,n}(x)
p_{k,n}(y),
\]
where $(p_{k,n})_k$ is the sequence of orthonormal polynomials with
respect to the weight function $e^{-nV(x)}$ on the real line. As $n
\to \infty$ the empirical eigenvalue distributions have an a.s.\
weak limit\footnote{i.e., for any bounded continuous function $f$, we have
$\lim\limits_{n\to \infty}\frac{1}{n} \sum_{j=1}^n f(x_j) = \int f d\mu^*$
almost surely.} $\frac{1}{n} \sum_{j=1}^n \delta_{x_j} \to \mu^*$
where $\mu^*$ is a non-random probability measure that is
characterized as the minimizer of the energy functional (Coulomb gas
picture)
\begin{equation} \label{energyOneMM}
    E_V(\mu) = \iint \log \frac{1}{|x-y|} d\mu(x) d\mu(y) + \int V(x) d\mu(x)
    \end{equation}
when taken over all probability measures on the real line. For a polynomial $V$ the minimizer $\mu^*$
is supported on a finite union of intervals \cite{DKMc}. In addition there
is a polynomial $Q$ of degree $\deg V -2$ such that
\[ \xi(z) = V'(z) - \int \frac{d\mu_1^*(s)}{z-s} \]
is the solution of a quadratic equation
\begin{equation} \label{speccurve1MM}
    \xi^2 - V'(z) \xi + Q(z) = 0.
    \end{equation}
From this it follows that $\mu_1^*$ has a density with respect to Lebesgue
measure that is real analytic in the interior of any of the intervals and that can
be written as
\[ \rho(x) = \frac{d\mu^*(x)}{dx} = \frac{1}{\pi} \sqrt{q^-(x)}, \qquad x \in \mathbb R \]
where $q^-$ denotes the negative part of the polynomial
\[ q(x) = \left( \frac{V'(x)}{2} \right)^2 - Q(x). \]

\section{Limiting eigenvalue distribution}
\subsection{Vector equilibrium problem}
Guionnet \cite{Gui} showed that the eigenvalues of the matrices $M_1$ and $M_2$
in the two matrix model \eqref{TwoMM} have a limiting distribution as $n \to \infty$.
The results of \cite{Gui} are in fact valid for a much greater class of random matrix models. The limiting
distribution is characterized as the minimizer of a certain functional, which is however
very different from the energy functional \eqref{energyOneMM} for the one matrix model.

Our aim is to develop an analogue of the Coulomb gas picture for the eigenvalues
of the matrices in the two matrix model \eqref{TwoMM}. We have been successful in doing
this for the eigenvalues of $M_1$ in the case of even polynomial potentials $V$ and $W$
with $W$ of degree $4$. Thus our assumptions are
\begin{itemize}
\item $V$ is an even polynomial with positive leading coefficient,
\item $W(y) = \frac{1}{4} y^4 + \frac{\alpha}{2} y^2$ with $\alpha \in \mathbb R$,
\item and $\tau > 0$ (without loss of generality).
\end{itemize}

We use the following notions from logarithmic potential theory \cite{ST}
\begin{align*}
    I(\mu,\nu) & = \iint \log \frac{1}{|x-y|} \, d\mu(x) d\nu(y),
    \qquad I(\mu) = I(\mu,\mu),
    \end{align*}
which define the mutual logarithmic energy $I(\mu, \nu)$
of two measures $\mu$ and $\nu$, and the logarithmic energy $I(\mu)$ of a measure $\mu$. Then the limiting
mean distribution of the eigenvalues of $M_1$ is characterized
by a vector equilibrium problem for three measures.
This involves an energy functional
\begin{multline} \label{energy2MM}
    E(\mu_1, \mu_2, \mu_3) =
    I(\mu_1) + I(\mu_2) + I(\mu_3)  \\ - I(\mu_1,\mu_2) - I(\mu_2,\mu_3)
        + \int V_1(x) \, d\mu_1(x) + \int V_3(x) d\mu_3(x)
\end{multline}
defined on three measures $\mu_1$, $\mu_2$, $\mu_3$.
Note that there is an attraction between the measures $\mu_1$ and $\mu_2$
and between the measures $\mu_2$ and $\mu_3$, while there is no direct
interaction between the measures $\mu_1$ and $\mu_3$. This type of interaction
is characteristic for a Nikishin system \cite{NS}.

The energy functional \eqref{energy2MM} depends on the external fields $V_1$ and $V_3$
that act on the measures $\mu_1$ and $\mu_3$ in \eqref{energy2MM}.
The vector equilibrium problem will also have an upper constraint $\sigma_2$
for the measure $\mu_2$. These input data take a very special form that we describe next.

\paragraph{External field $V_1$:} The external field that acts on $\mu_1$
is defined by
\begin{equation} \label{defV1}
    V_1(x) = V(x) + \min_{s \in \mathbb R} (W(s) - \tau xs),
    \end{equation}
where we recall that $W(s) = \frac{1}{4} s^4 + \frac{\alpha}{2} s^2$.
For the case $\alpha = 0$, this is simply $V_1(x) = V(x) - \frac{3}{4} |\tau x|^{4/3}$.

\paragraph{External field $V_3$:}
The external field that acts on the third measure is absent if $\alpha \geq 0$, i.e.,
\[ V_3(x) \equiv 0 \qquad \text{if } \alpha \geq 0. \]

The function $s \in \mathbb R \mapsto W(s) - \tau x s$ has a global
minimum at $s = s_1(x)$ and this value plays a role in the
definition of $V_1$, see \eqref{defV1}. For $\alpha < 0$, and $x \in
(-x^*(\alpha), x^*(\alpha))$, where
\[ x^*(\alpha) = \frac{2}{\tau} \left( \frac{-\alpha}{3} \right)^{3/2}, \qquad \alpha < 0. \]
The function $s \in \mathbb R \mapsto W(s) - \tau xs$  has another
local minimum at $s = s_2(x)$, and a local maximum at $s= s_3(x)$.

Then $V_3$ is defined by
\begin{equation} \label{defV3}
    V_3(x) = \left(W(s_3(x)) - \tau x s_3(x) \right) - \left(W(s_2(x)) - \tau x s_2(x)
    \right)
    \end{equation}
if $x \in (-x^*(\alpha), x^*(\alpha))$, and
\[ V_3(x) \equiv 0 \qquad \text{otherwise}. \]

\paragraph{Upper constraint $\sigma_2$:} The upper constraint $\sigma_2$ that
acts on the second measure is the measure on the imaginary axis with the density
\begin{equation} \label{defsigma2}
    \frac{d\sigma_2(z)}{|dz|} = \frac{\tau}{\pi} \max_{s^3 + \alpha s = \tau z} \Re s, \qquad z \in i \mathbb R.
    \end{equation}
In case $\alpha = 0$ this simplifies to $\frac{d \sigma_2}{|dz|} = \frac{\sqrt{3}}{2\pi} \tau^{4/3} |z|^{1/3}$.

If $\alpha < 0$ then the density of $\sigma_2$ is positive and real analytic on the full imaginary
axis. If $\alpha > 0$ then the support of $\sigma_2$ has a gap around $0$:
\[ \supp(\sigma_2) = (-i\infty, -i y^*(\alpha)] \cup [i y^*(\alpha), i \infty), \]
where
\[ y^*(\alpha) = \frac{2}{\tau} \left( \frac{\alpha}{3} \right)^{3/2}, \qquad   \alpha > 0. \]

The following result is Theorems 1.1 in \cite{DKM}.
\begin{theorem} \label{mainTHM1}
There is a unique minimizer $(\mu_1^*, \mu_2^*, \mu_3^*)$  of the energy functional \eqref{energy2MM}
subject to the conditions
\begin{enumerate}
\item[\rm (a)] $\mu_1$ is a measure on $\mathbb R$ with $\mu_1(\mathbb R ) = 1$,
\item[\rm (b)] $\mu_2$ is a measure on $i \mathbb R$ with $\mu_2( i \mathbb R ) = 2/3$,
\item[\rm (c)] $\mu_3$ is a measure on $\mathbb R$ with $\mu_3(\mathbb R ) = 1/3$,
\item[\rm (d)] $\mu_2 \leq \sigma_2$,
\end{enumerate}
with input data $V_1$, $V_3$, and $\sigma$ as described above.
\end{theorem}

The proof of the existence of a minimizer was completed and
simplified in \cite{HK}, see subsection \ref{subsec:energyHK} below.

Now that we have existence and uniqueness, it is natural to ask
about further properties of the minimizer. The three measures
$\mu_1^*$, $\sigma-\mu_2^*$ and $\mu_3^*$ are absolutely continuous
with respect to the Lebesgue measure with densities that are real
analytic in the interior of their supports, except possibly at the
origin. Furthermore, denoting by $S(\mu)$ the support of a measure
$\mu$, we have
\begin{itemize}
\item the support of $\mu^*_1$ is a finite union of bounded intervals on the real line;
\item there exist $c_2 \geq 0$ such that $S(\sigma_2 - \mu^*_2) = i\mathbb R \setminus (-ic_2, ic_2)$, and
if $c_2 > 0$ then the density of $\sigma_2 - \mu_2^*$ vanishes like a square root at $\pm ic_2$;
\item there exist $c_3 \geq 0$ such that $S(\mu^*_3) = \mathbb R \setminus (-c_3, c_3)$, and
if $c_3 > 0$ then the density of $\mu_3^*$ vanishes like a square root at $\pm c_3$.
\end{itemize}

In a generic situation, the density of $\mu_1^*$ is strictly positive
in the interior of its support and vanishes like a square root at endpoints.
In addition strict inequality holds in the variational inequality outside the support $S(\mu_1^*)$.
Moreover, generically if $c_2 = 0$ the density of  $\sigma-\mu_2^*$ is positive at the origin,
and likewise if $c_3 = 0$ the density of $\mu_3^*$ is positive at the origin. If we are in such a generic situation,
then we say that $(V,W,\tau)$ is \emph{regular}. See \cite[section 1.5]{DKM} for more details and a discussion on the singular situations that may occur.

The following is Theorem 1.4 in \cite{DKM}.

\begin{theorem} \label{mainTHM2}
Let $\mu_1^*$ be the first component of the minimizer in Theorem 1,
and assume that  $(V,W, \tau)$ is regular, then as $n \to \infty$
with $n \equiv 0 \, (\mathrm{mod} 3)$, the mean eigenvalue
distribution of $M_1$ convergences to $\mu_1^*$.
\end{theorem}

We are convinced that the theorem is also valid in the singular cases,
which correspond to phase transitions in the two matrix model. The condition that $n$ is a multiple of three is
non-essential as well. It is imposed for convenience in the analysis.

In \cite{DKM} only the convergence of mean eigenvalue distributions was considered, which is a rather weak
form of convergence. However, when combined with the results of \cite{Gui} it will actually follow that
the empirical eigenvalue distributions of $M_1$ tend to $\mu_1^*$ almost surely.

The analysis of \cite{DKM} also proves the usual universality results for local eigenvalue statistics
in Hermitian matrix ensembles, given by the sine kernel in the bulk of the spectrum and by the Airy kernel
at edge points. In non-regular situations one may find Pearcey and Painlev\'e II kernels, while in multi-critical
cases new kernels may appear. This was indeed proved recently in \cite{DG}, see subsection
\ref{subsec:criticalDG} below.

\subsection{Riemann surface}

A major ingredient in the asymptotic analysis in \cite{DKM} is the construction of an appropriate Riemann surface (or spectral curve), which plays a role similar to the algebraic equation \eqref{speccurve1MM}
in the one matrix model. The existence of such a Riemann surface is implied by
the work of Eynard \cite{Eyn} on the formal two matrix model. Our approach is different
from the one of Eynard, in that we use the vector equilibrium problem to construct the
Riemann surface, and in a next step we define a meromorphic function on it.

The main point is that the supports  $S(\mu_1^*)$, $S(\sigma-\mu_2^*)$ and $S(\mu_3^*)$  associated to the minimizer in Theorem 1, determine the cut structure of a Riemann surface
\[ \mathcal R = \bigcup_{j=1}^4 \mathcal R_j \]
with four sheets
\begin{align*}
    \mathcal R_1 & = \overline{\mathbb C} \setminus S(\mu^*_1), \\
    \mathcal R_2 & = \mathbb C \setminus \left(S(\mu^*_1) \cup S(\sigma_2 -\mu^*_2)\right), \\
    \mathcal R_3 & = \mathbb C \setminus \left(S(\sigma_2-\mu^*_2) \cup S(\mu^*_3)\right), \\
    \mathcal R_4 & = \mathbb C \setminus S(\mu^*_3).
\end{align*}
The sheet $\mathcal R_j$ is glued to the next sheet $\mathcal R_{j+1}$ along the
common cut in the usual crosswise manner.
The meromorphic function on $\mathcal R$ arises in the following way, see Proposition 4.8 of
\cite{DKM}.

\begin{proposition} \label{propMero}
The function
\[  \xi_1(z) = V'(z) - \int \frac{d\mu^*_1(x)}{z-x}, \qquad z \in \mathcal R_1, \]
extends to a meromorphic function on the  Riemann surface $\mathcal R$ whose only poles are
at infinity. There is a pole of order $\deg V$ at infinity on the first sheet, and a simple
pole at the other point at infinity.
\end{proposition}

The proof of Proposition \ref{propMero} follows from the
Euler-Lagrange variational conditions that are associated with the
vector equilibrium problem. See section 4.2 of \cite{DKM} for
explicit expressions for the meromorphic continuation of $\xi_1$ to
the other sheets.

It follows from Proposition \ref{propMero} that $\xi_1$ is one of
the  solutions of a quartic equation, which is the analogue of the
quadratic equation \eqref{speccurve1MM} that is relevant in the
one matrix model.

\section{About the proof}

We describe the main tools that are used in the proof of Theorem
\ref{mainTHM2}.


\subsection{Biorthogonal polynomials}

We  make use of the integrable structure of the two matrix model
that is described in terms of biorthogonal polynomials. In this
context the biorthogonal polynomials are two sequences of
monic polynomials $(p_{j,n})_j$ and $(q_{k,n})_k$
(depending on $n$) with $\deg p_{j,n} = j$ and $\deg q_{k,n} = k$, that satisfy
\[
    \int_{-\infty}^{\infty} \int_{-\infty}^{\infty} p_{j,n}(x) q_{k,n}(y)
    e^{-n(V(x) + W(y) - \tau xy)} dx dy = h_{k,n} \delta_{j,k},
     \]
see \cite{Ber, BEH1, BEH2, EMc, EyMe}. These polynomials uniquely
exist, have real and simple zeros \cite{EMc}, and in addition the
zeros of $p_{j,n}$ and $p_{j+1,n}$ interlace, as well as those of
$q_{k,n}$ and $q_{k+1,n}$, see \cite{DGK}.

There is an explicit expression for the joint p.d.f.\ of the eigenvalues of $M_1$ and $M_2$
\begin{equation} \label{pdf2MM}
    \frac{1}{(n!)^2} \det \begin{pmatrix}  K_n^{(1,1)}(x_i,x_j)  & K_n^{(1,2)}(x_i,y_j) \\
                K_n^{(2,1)}(y_1,y_j) & K_n^{(2,2)}(y_i,y_j) \end{pmatrix}
                \end{equation}
with $4$ kernels that are expressed in terms of the biorthogonal polynomials
and their transformed functions
\begin{align*}
    Q_{k,n}(x) &= \int_{-\infty}^{\infty} q_{k,n}(y) e^{-n\left(V(x) + W(y)-\tau x y \right)}  dy,\\
    P_{j,n}(y) &= \int_{-\infty}^{\infty} p_{j,n}(x) e^{-n\left(V(x) + W(y) -\tau x y \right)}  dx,
\end{align*}
as follows
\begin{align} \label{Kn11}
K_n^{(1,1)}(x_1,x_2) & = \sum_{k=0}^{n-1} \frac{1}{h_{k,n}^2} p_{k,n}(x_1) Q_{k,n}(x_2), \\
K_n^{(1,2)}(x,y) & =\sum_{k=0}^{n-1} \frac{1}{h_{k,n}^2} p_{k,n}(x) q_{k,n}(y),  \nonumber \\
K_n^{(2,1)}(y,x) & =\sum_{k=0}^{n-1} \frac{1}{h_{k,n}^2} P_{k,n}(y) Q_{k,n}(x) -e^{-n\left(V(x)+W(y)-\tau x y\right)}, \nonumber \\
K_n^{(2,2)}(y_1,y_2) & =\sum_{k=0}^{n-1} \frac{1}{h_{k,n}^2} P_{k,n}(y_1) q_{k,n}(y_2). \nonumber
\end{align}
The joint p.d.f.\ \eqref{pdf2MM} is determinantal, which means that
eigenvalue correlation functions have determinantal expressions with
the same kernels $K_n^{(i,j)}$, $i,j=1,2$. In particular, after
averaging out the eigenvalues of $M_2$ we get that the eigenvalues
of $M_1$ are a determinantal point process with kernel
$K_n^{(1,1)}$.

A natural first step to compute the asymptotic behavior of the polynomials and hence the kernels, is to formulate a Riemann-Hilbert problem   ($\equiv$ RH problem) for the polynomials. Several different formulations exist in the literature \cite{BEH2,EMc,Kapaev,KMc}. The analysis in  \cite{DK,DKM,Mo} is based on the RH problem in \cite{KMc} that we will discuss in the next subsection.
\subsection{Riemann-Hilbert problem}
It turns out that the kernel \eqref{Kn11} has a special structure
which relates it to multiple orthogonal polynomials and the
eigenvalues of $M_1$ (after averaging over $M_2$) are an example of
a multiple orthogonal polynomial ensemble \cite{Kui}. This is due to
the following observation of Kuijlaars and McLaughlin \cite{KMc}.
\begin{proposition}
Suppose $W$ is a polynomial of degree $r+1$, and let
\[  w_{k,n}(x) = \int_{-\infty}^{\infty} y^k e^{-n (V(x) + W(y) - \tau xy)} dy,
    \quad k=0, \ldots, r-1. \]
Then the biorthogonal polynomial $p_{j,n}$ satisfies
\begin{equation} \label{MOPcondition}
    \int_{-\infty}^{\infty} p_{j,n}(x) x^l w_{k,n}(x) \, dx = 0,
    \quad l = 0, \ldots, \left\lceil \frac{j-k}{r} \right\rceil -1, \end{equation}
    for $k=0, \ldots, r-1$.
    \end{proposition}
The conditions \eqref{MOPcondition} are known as multiple orthogonality conditions \cite{Apt},
and they characterize the biorthogonal polynomials.

The advantage of the formulation as multiple orthogonality is that these
polynomials are characterized by a RH problem of size $(r+1) \times (r+1)$, \cite{VAGK},
which we state here for the case $r = 3$ and for $j = n$ with $n$ a multiple of three.
Then the RH problem has size $4 \times 4$ and it asks for a $4 \times 4$ matrix valued
function $Y$ on $\mathbb C \setminus \mathbb R$ such that
\begin{enumerate}
\item[(a)] $Y : \mathbb C \setminus \mathbb R \to \mathbb C^{4 \times 4}$ is analytic,
\item[(b)] $Y_+(x) = Y_-(x) \begin{pmatrix} 1 & w_{0,n}(x) & w_{1,n}(x) & w_{2,n}(x) \\
    0 & 1 & 0 & 0 \\ 0 & 0 & 1 & 0 \\ 0 & 0 & 0 & 1 \end{pmatrix}$ \, for $x \in \mathbb R$, where
    $Y_+(x)$ ($Y_-(x))$) denotes the limiting value of $Y(z)$ as $z \to x$ from the upper (lower) half plane,
    \item[(c)] $Y(z) = \left(I_4 + O\left(\frac{1}{z}\right)\right)
        \begin{pmatrix} z^{n} & 0 & 0 & 0 \\ 0 & z^{-n/3} & 0 & 0 \\ 0 & 0 & z^{-n/3} & 0 \\ 0 & 0 & 0 & z^{-n/3}
        \end{pmatrix} $ \, as $z \to \infty$.
    \end{enumerate}

The RH problem has a unique solution which is given by
\[ Y =
    \begin{pmatrix} p_{n,n} & C(p_{n,n} w_{0,n}) & C(p_{n,n} w_{1,n}) & C(p_{n,n} w_{2,n}) \\
    p_{n,n}^{(0)} & C(p_{n,n}^{(0)} w_{0,n}) & C(p_{n,n}^{(0)} w_{1,n}) & C(p_{n,n}^{(0)} w_{2,n}) \\
    p_{n,n}^{(1)} & C(p_{n,n}^{(1)} w_{0,n}) & C(p_{n,n}^{(1)} w_{1,n}) & C(p_{n,n}^{(1)} w_{2,n}) \\
    p_{n,n}^{(2)} & C(p_{n,n}^{(2)} w_{0,n}) & C(p_{n,n}^{(2)} w_{1,n}) & C(p_{n,n}^{(2)} w_{2,n})
\end{pmatrix},
\]
where $p_{n,n}$ is the $n$th degree biorthogonal polynomial,
$p_{n,n}^{(0)}$, $p_{n,n}^{(1)}$, $p_{n,n}^{(2)}$ are three
polynomials of degree $\leq n-1$ that satisfy certain multiple
orthogonal conditions and $Cf$ is the Cauchy transform
\[ Cf (z) = \frac{1}{2\pi i} \int_{-\infty}^{\infty} \frac{f(x)}{x-z} dx. \]

By using the the Christoffel-Darboux formula for multiple orthogonal
polynomials \cite{BK1,DaK} the correlation kernel $K_n^{(1,1)}$ for the
eigenvalues of $M_1$ can be expressed in terms of the solution of
the RH problem as follows
    \begin{multline} \label{kernelKnandY} K_n^{(1,1)}(x,y) =
     \begin{pmatrix} 0 & w_{0,n}(y) & w_{1,n}(y) & w_{2,n}(y) \end{pmatrix}
            \frac{Y_+^{-1}(y) Y_+(x)}{2\pi i(x-y)} \begin{pmatrix} 1 \\ 0 \\ 0 \\ 0
            \end{pmatrix}.
            \end{multline}
Multiple orthogonal polynomials and RH problems are also used for random
matrices with external source \cite{BDK,BK1} and models of non-intersecting
paths \cite{KMW}. In these cases, correlation kernels
for the relevant statistical quantities are also expressed in terms
of the corresponding RH problem through \eqref{kernelKnandY}.

\subsection{Steepest descent analysis}

The remaining part of the proof of Theorem \ref{mainTHM2} is an asymptotic analysis
of the RH problem via an extension of the Deift-Zhou steepest descent method \cite{DKMVZ,DZ}.
The vector equilibrium problem and the Riemann surface play a crucial role
in the transformations in this analysis. For the precise transformations
and the many details that are involved we refer the reader to \cite{DKM}.
Following the effect of the transformations on the kernel \eqref{kernelKnandY}, one finds that
\[ \lim_{n \to \infty} \frac{1}{n} K_n^{(1,1)}(x,x) = \frac{d\mu_1^*(x)}{dx}, \]
which is what is needed to establish the theorem.

A  somewhat similar steepest descent analysis is done in \cite{BDK}
for a random matrix model with external source, where vector
equilibrium problems and Riemann surfaces also play an important
role.

\section{Further developments}
\subsection{Critical behavior in the quadratic/quartic model} \label{subsec:criticalDG}

For the case $V(x) = \frac{1}{2} x^2$ the spectral curve can be
computed and a classification of all possible cases can be made
explicitly.
\begin{description}
\item[Case I:] $0 \in S(\mu_1^*) \cap S(\mu_3^*)$ and
$0 \not \in S(\sigma_2 - \mu_2^*)$;
\item[Case II:] $0 \in S(\mu_3^*)$ and $0 \not\in S(\mu_1^*) \cup S(\sigma_2 - \mu_2^*)$;
\item[Case III:] $0 \in S(\sigma_2 - \mu_2^*)$ and $0 \not\in S(\mu_1^*) \cup  S(\mu_3^*)$;
\item[Case IV:] $0 \in S(\mu_1^*)$ and $0 \not\in S(\sigma_2 - \mu_2^*) \cup S(\mu_3^*)$.
\end{description}
Phase transitions between the regular cases represent the critical
cases.
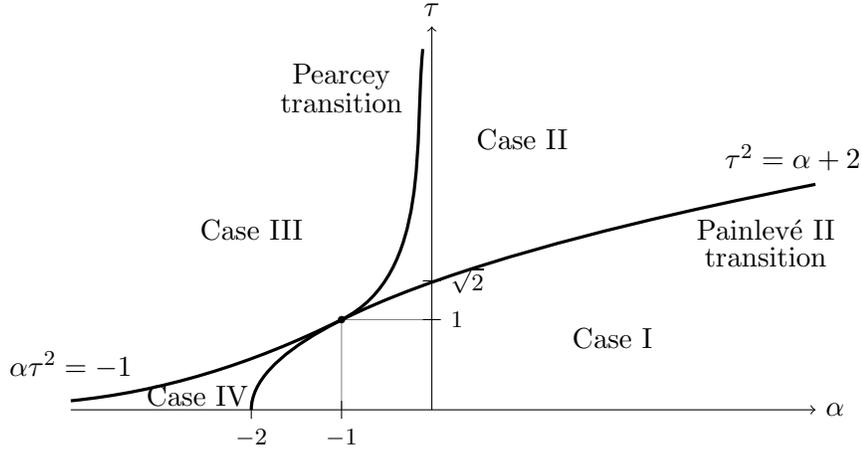
\begin{figure}[t] 
\begin{tikzpicture}[scale=1.2]
\draw[->](0,0)--(0,4.25) node[above]{$\tau$};
\draw[->](-4,0)--(4.25,0) node[right]{$\alpha$};
\draw[help lines] (-1,0)--(-1,1)--(0,1);
\draw[black, very thick,rotate around={-90:(-2,0)}] (-2,0) parabola (-4.5,6.25);
\draw[black, very thick] (3.7,2) node{Painlev\'e II};
\draw[black] (3.7,1.7) node{transition};
\draw[black, very thick] (4,2.8) node{$\tau^2= \alpha+2$};
\draw[black, very thick] (-1,1)..controls (0,1.5) and (-0.2,3).. (-0.1,4)
             (-1,1)..controls (-2,0.5) and (-3,0.2).. (-4,0.1);
\draw[black, very thick] (-1,3.7) node{Pearcey};
\draw[black] (-1,3.4) node{transition};
\draw[black, very thick] (-4,0.5) node{$\alpha \tau^2=-1$};
\filldraw  (-1,1) circle (1pt);
\draw (0.1,1) node[font=\footnotesize,right]{$1$}--(-0.1,1);
\draw (-1,0.1)--(-1,-0.1) node[font=\footnotesize,below]{$-1$};
\draw (-2,0.1)--(-2,-0.1) node[font=\footnotesize,below]{$-2$};
\draw (0.1,1.43) node [font=\footnotesize,right]{$\sqrt 2$}--(-0.1,1.43);
\draw[very thick] (2,0.8) node[fill=white]{Case I}
                  (-2.6,0.15) node{Case IV} 
                  (-2,2) node[fill=white]{Case III}
                  (1,3) node[fill=white]{Case II}; 
\end{tikzpicture}
\caption{Phase diagram for the quadratic case $V(x) = \frac{1}{2} x^2$. \label{fig:phasediagram}}
\end{figure}

The quadratic/quartic model depends on two parameters, namely the
coupling constant $\tau$ and the number $\alpha$ in the quartic
potential $W(y) = \frac{y^4}{4} + \alpha \frac{y^2}{2}$. Figure
\ref{fig:phasediagram} (taken from \cite{DGK}) shows the phase
diagram in the $\alpha$-$\tau$ plane. Critical behavior takes place
on the curves $\tau^2 = \alpha + 2$ and $\alpha \tau^2 = -1$. On the
parabola $\tau^2 = \alpha + 2$ a gap appears around $0$ in the
support of either $\mu_1^*$ (if one moves from Case I to Case II) or
$\mu_3^*$ (if one moves from Case I to Case IV). This is a
transition of Painlev\'e II type which also appears in the opening
of gaps in one matrix models \cite{BI,CK}. On the curve $\alpha
\tau^2 = -1$ a gap appears in the support of either $\mu_1^*$  (if one moves from Case IV to Case III) or  $\mu_3^*$ (if one moves from Case II to Case III),
while simultaneously the gap in the support of $\sigma_2 - \mu_2^*$
closes. This is a transition of Pearcey type, which was observed
before in the random matrix model with external source and in the
model of non-intersecting Brownian motions \cite{BK2,BH,TW}.

The phase diagram has a very special point $\alpha = -1$, $\tau = 1$
which is on both critical curves, and where all four regular cases
come together. For these special values, the density of $\mu_1^*$
vanishes like a square root at the origin, which is an interior
point of $S(\mu_1^*)$. The local analysis at this point was done
very recently by Duits and Geudens \cite{DG}. They found that in the
asymptotic limit, the local eigenvalue correlation kernels around
$0$ are closely related to the limiting kernels that describe the
tacnode behavior for non-intersecting Brownian motions \cite{AFM,
DKZ, Joh}. More precisely, the kernels can be expressed in terms of an extension of the same $4 \times 4$ RH
problem in \cite{DKZ}. However, they are constructed in a different way out of this $4\times 4$ RH problem and, as a result, these kernels are not the same.
\subsection{Vector equilibrium problems} \label{subsec:energyHK}
The analysis in \cite{DKM} of the vector equilibrium problem
was not fully complete, since the lower semi-continuity of the energy
functional \eqref{energy2MM} was implicitly assumed but not established in \cite{DKM}.

In the recent papers \cite{BKMW, HK} the vector equilibrium problem was studied in
a more systematic way, in the more general context of an energy functional
for $n$ measures
\begin{equation} \label{energyfunctionalHK}
    E(\mu_1, \ldots, \mu_n)
    = \sum_{i=1}^n \sum_{j=1}^n c_{ij} I(\mu_i, \mu_j) + \sum_{j=1}^n \int V_j(x)
    d\mu_j(x),
    \end{equation}
where $C = (c_{ij})_{i,j=1}^n$ is a real symmetric positive definite
matrix (in \cite{BKMW} also semidefinite interaction matrices are
considered). The external fields  $V_j : \Sigma_j \to \mathbb R \cup
\{\infty\}$ are lower semi-continuous with domains $\Sigma_j$ that
are closed subsets of $\mathbb C$. Let $m_1, \ldots, m_n$ be given
positive numbers and assume that for every $i = 1, \ldots, n$,
\[ \liminf_{|x| \to \infty} \left( V_i(x) - \left( \sum_{j=1}^n c_{ij} m_j \right) \log(1 + |x|^2) \right) > - \infty. \]
Under these assumptions it is shown in \cite{HK} that the energy
functional \eqref{energyfunctionalHK}, restricted to the set  of
measures with $\mu_j(\Sigma_j) = m_j$ for $j=1, \ldots, n$,
\begin{itemize}
\item[(a)] has  compact sub-level sets $\{ E \leq  \alpha \}$ for every $\alpha \in \mathbb R$,
(so $E$ is in particular lower semi-continuous), and
\item[(b)] is strictly convex on the subset where it is finite.
\end{itemize}
This guarantees existence and uniqueness of a minimizer of
\eqref{energyfunctionalHK}, provided that $E$ is not identically
infinite. Existence and uniqueness of a minimizer readily extends to
situations where the domain of $E$ is further restricted by upper
constraints $\mu_j \leq \sigma_j$ for $j=1, \ldots, n$, again
provided that $E$ is not identically infinite on this domain. In
particular, this applies to the energy functional \eqref{energy2MM}
for the two matrix model with quartic potential with the constraint
$\mu_2 \leq \sigma_2$ described in section 2.

\subsection{Open problems}

There are numerous intriguing questions and open problems that arise out of our analysis.

\begin{enumerate}
\item[(a)] What is the motivation for the central vector equilibrium problem?
In the one-matrix model there is a direct way to come from the joint eigenvalue probability
density to the equilibrium problem. We do not have this direct link for the two matrix model.
\item[(b)] How is the vector equilibrium problem related to the variational problem from \cite{Gui}?
\item[(c)] A possibly related question: is there a large deviation principle associated with the
vector equilibrium problem? See e.g.\ \cite{AGZ} for the large deviations interpretation
of the equilibrium problem for the one matrix model.
\item[(d)] Our analysis is restricted to even potentials $V$ and $W$. This restriction
provides a symmetry of the problem around zero, which is the reason why the second measure $\mu_2$ in
the vector equilibrium problem is supported on the imaginary axis. If we remove the symmetry
then probably we would have to look for a contour that replaces the imaginary axis.
It is likely that such a contour would be an $S$-curve in a certain external field, but
at this moment we do not know how to handle this situation. See \cite{MaRa,Rak} for important
recent developments around $S$-curves for scalar equilibrium problems.
\item[(e)] Extensions to higher degree $W$ is wide open. If $\deg W = d$ then one would
expect a vector equilibrium problem for $d-1$ measures.  It may be that $S$-curves are needed
for $d \geq 6$, even in the case of even potentials.
\item[(f)] Exploration of further critical phenomenon in the two matrix model.
\end{enumerate}

\end{document}